# Designing Extremely Low-Power Topological Transistors with 1T′-MoS$_2$ and HZO for Cryogenic Applications

*Yosep Park[a], Yungyeong Park [b], Hyeonseok Choi[b], Subeen Lim[b], and Yeonghun Lee[a,b,c]\**

[a]Department of Intelligent Semiconductor Engineering, Incheon National University, 119 Academy-ro, Yeonsu-gu, Incheon 22012, Republic of Korea

[b]Department of Electronics Engineering, Incheon National University, Incheon National University, 119 Academy-ro, Yeonsu-gu, Incheon 22012, Republic of Korea

[c]Research Institute for Engineering and Technology, Incheon National University, 119 Academy-ro, Yeonsu-gu, Incheon 22012, Republic of Korea

*Email: y.lee@inu.ac.kr

**ABSTRACT :** Large-scale quantum computing requires cryogenic electronic controllers such as control/readout circuit and routing circuit. However, current technologies face high power dissipation problems, hindering large-scale qubit integration. Here, we theoretically propose extremely low-power cryogenic topological transistors, i.e., negative-capacitance topological insulator field-effect transistors (NC-TIFETs). By combining a gate-field-induced two-dimensional 1T′-Molybdenum Disulfide (MoS$_2$) topological channel with a Hafnium–Zirconium Oxide (HZO) ferroelectric gate insulator, NC-TIFETs exhibit an extremely steep-slope transfer curve and ultra-high transconductance at low drain voltage ($V_D$). Therefore, NC-TIFETs are the compelling candidate for minimizing power dissipation in the cryogenic electronic interfaces essential for large-scale quantum computing systems.

Researchers have been developed quantum computers to solve complex problems that classical computers cannot solve[1–3]. To enable large-scale quantum computing systems, electronic controllers (i.e., control/readout circuits and routing circuits) that can minimize the number of coaxial cables to operating quantum computing systems (Figure 1a)[4–8]. However, as qubit scalability increases, these controllers are faced with the challenge of consuming power that exceeds the cooling power limits available in conventional dilution refrigerators[3,9,10]. In order to resolve the cooling power problems, III-V high-electron-mobility transistors (HEMTs) have been investigated for their potential of low-power dissipation and high-frequency operation[7,8,11–13]. Nevertheless, the power dissipation of HEMTs has remained too high for operating large-scale quantum computing systems[9,10,14]. To overcome the power consumption bottleneck in electronic controllers, we propose extremely low-power transistors based on a novel operational principle, utilizing two-dimensional topological insulators (2D TIs). 2D TIs offer two key advantages: their topologically protected edge states enable ballistic transport by suppressing backscattering, and they undergo a topological phase transition from the quantum spin Hall (QSH) to the normal insulator (NI) phase as an out-of-plane electric field ($E_z$) breaks inversion symmetry[15,16]. The properties of these 2D TIs make them applicable as TIFETs, characterized by novel switching mechanisms and high on-state currents.

In addition to the aforementioned properties, the TIFET possesses a desirable character that can maximize the gain of NC operation that ferroelectric materials introduce. In conventional FETs, the quantum capacitance ($C_Q$) formed at the onset of inversion is known to cause undesirable hysteresis in the $I_D$–$V_G$ characteristics, thereby limiting the gain from the NC effect[17]. In contrast, during the switching operation of TIFETs, only the edge states are modulated while there are no charges in most part (i.e., bulk) of the TIs. Accordingly, such inversion in FETs does not occur during the TIFET operation, which allows us to take full advantage of the NC to amplify the electric field in the TIFETs. Although the conceptual framework of NC-TIFET has been proposed previously[18], our work provides comprehensive analyses including current-voltage characteristics and discusses superior performance in the context of large-scale quantum computers.

In this work, we theoretically propose a device model for the NC-TIFET by combining the tight-binding (TB) model[19], $\mathbf{k} \cdot \mathbf{p}$ model[20,21], non-equilibrium Green's function (NEGF) formalisms[22,23], and incorporating the NC effects via Landau–Khalatnikov (L–K) equation[24]. Our model provides a theoretical  analysis for achieving steep-slope switching and maximizing the $E_z$ amplification through the incorporation of ferroelectric materials. Subsequently, the voltage gain ($A_V$) of the NC-TIFET is compared to that of a conventional negative capacitance field-effect transistor (NCFET), analyzing the differences in $A_V$ in the context of their equivalent capacitance circuits and charge modulation characteristics. Finally, the simulation results demonstrated that NC-TIFETs exhibit an extremely steep-slope transfer curve compared with typical TIFETs. Additionally, the steep-slope transfer curve results in ultra-high transconductance ($g_m$) at low $V_D$, thereby maximizing its potential for cryogenic applications in quantum computing electronic controllers, such as control, readout, and routing circuits.

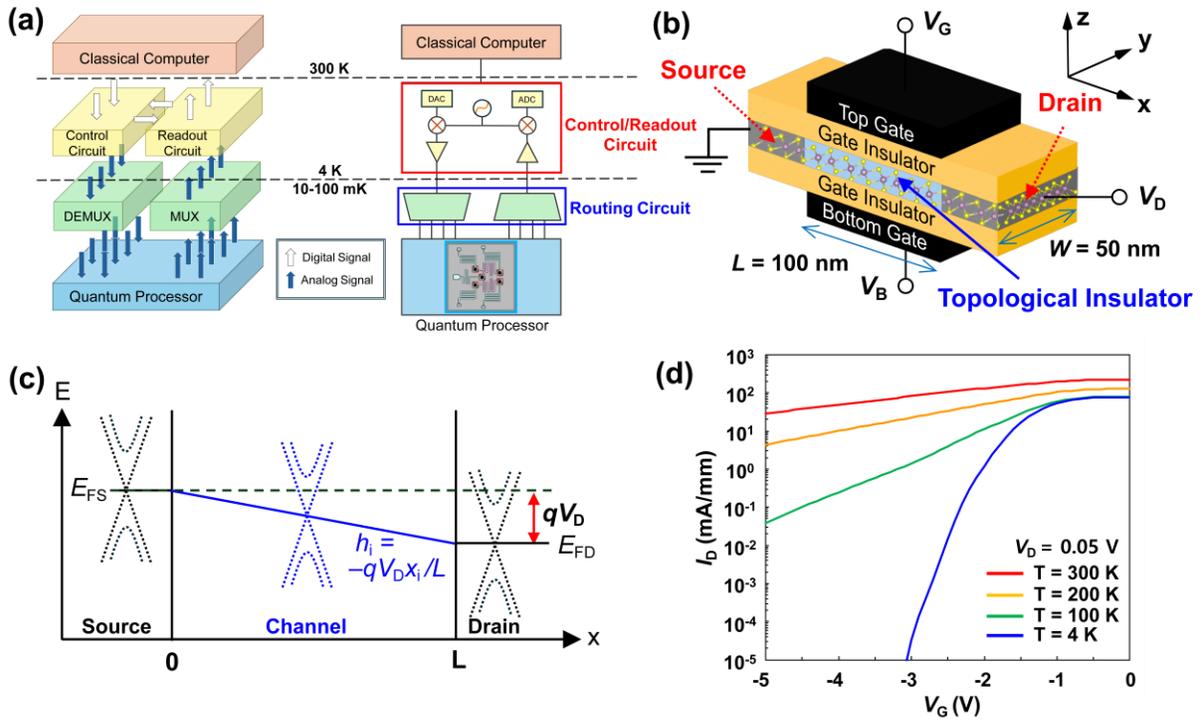

**Figure 1.** Conceptual architecture of a large-scale quantum computing system and TIFET device characteristics. (a) Schematic of large-scale quantum computing architecture and its distinct temperature stages. The quantum processor and routing circuits (MUX/DEMUX) operate at 10–100 mK, the control/readout circuits at 4 K, and the classical computer at room temperature (300 K). Digital and analog signal paths are indicated. (b) Device structure of double-gate TIFETs.

The channel consists of a 1T′-MoS$_2$ topological insulator (TI). Gate insulator consists of dielectric layer (in TIFETs) or ferroelectric materials (in NC-TIFETs). (c) The energy band diagram in real space along the channel direction with $E$-$k$ band structures for the on-state. $E_{FS}$ and $E_{FD}$ represent the Fermi level in the source and the drain, respectively, with an applied drain voltage $qV_D$. (d) Transfer characteristics ($I_D$–$V_G$) of the TIFET with a 1T′-MoS$_2$ channel, swept from $T$ = 4 K to 300 K at fixed drain voltage $V_D$ = 0.05 V.

The TIFET comprises a double-gate structure with a nanoribbon channel constructed by a topological insulator (TI), integrated with source and drain terminals. The TI channel is sandwiched by gate insulators—a dielectric layer for a TIFET or a ferroelectric layer for an NC-TIFET—and is controlled by top ($V_G$) and bottom ($V_B$) gates. (Figure 1b) The device has a channel length ($L$) of 100 nm and a channel width ($W$) of 50 nm, where parameters are optimized for avoiding tunneling leakage and finite size effect (Figure S1 of Supporting Information 5). We selected 1T′-MoS$_2$ as the channel material due to its low critical field ($E_c$)[20], which enables rapid topological phase transitions for ON-OFF modulation. The parameters used for the TIFET model are outlined in the Supporting Information 1 and 2.

For the TIFET to operate, the $E_z$ for ON–OFF state modulation and the $V_D$ for driving the current are essential. Figure 1c illustrates the energy band diagram of the TIFET under an applied $V_D$. Due to the applied $V_D$, the channel potential is assumed to vary linearly along the channel, creating a difference between the source and drain Fermi levels ($E_{FS}$ and $E_{FD}$, respectively). This potential difference, defined as $E_{FS} = E_{FD} + qV_D$, drives a drain current governed by quantum conductance. To implement this linear potential drop in our model, we add onsite energy term $h_i = -\frac{qV_D x_i}{L}$, where $q$ is elementary charge, $x_I$ is the position of site $i$ from the source to drain region. Figure 1d shows the $I_D$–$V_G$ of the TIFET at various temperatures ranging from 4 K to 300 K, with $V_D = 0.05$ V. Due to its ballistic edge-state transport, the TIFET enables a high on-state current governed by quantum conductance. At room temperature (300K), the device performance is severely degraded. The small bandgap of 1T′-MoS$_2$ (45–47 meV)[20,21], allows thermally excited carriers to create a leakage path, bypassing gate control and resulting in a poor on/off ratio of approximately 10 (at $V_G = -5$ V). At cryogenic temperatures, however, this thermal leakage path is strongly suppressed. As seen in Figure 1d, cooling the device to 4 K dramatically improves the on/off ratio, highlighting that while the TIFET is unsuitable for room-temperature logic, its high switching performance is maximized in cryogenic environments.

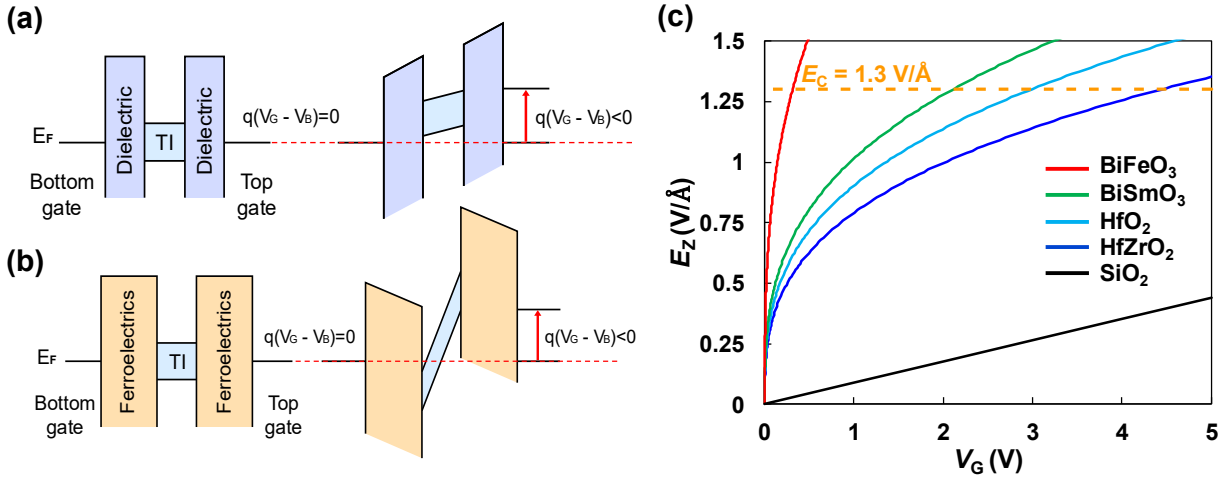

**Figure 2**. Mechanism of electric field amplification in NC-TIFETs. Energy band diagram of (a) a TIFET with a dielectric insulator layer and (b) an NC-TIFET with a ferroelectric insulator. The diagrams illustrate the state at zero gate bias ($q(V_G - V_B) = 0\ V$, left) and under an applied gate bias ($q(V_G - V_B) < 0\ V$), right), respectively. (c) The $E_z$ as a function of $V_G$ with a 1T′-MoS$_2$ channel, using various gate insulators : BiFeO$_3$[25], BiSmO$_3$[26], HfO$_2$[27], HfZrO$_2$[28], and SiO$_2$ (0.5 nm

thickness). The dashed orange line indicates the $E_c$ point, required to the topological phase transition.

The $E_z$ amplification mechanism of TIFETs, which depends on the gate insulator type, is illustrated in Figure 2a-b. When the gate bias is not applied, TIFETs maintain their topological phase, characterized by the presence of topological edge states. Conversely, when a gate bias is applied to TIFETs with a dielectric insulator layer, a limited $E_z$ is obtained due to the voltage drop across the insulator. In contrast, the NC-TIFET utilizes the NC effects of a ferroelectric layer to generate internal voltage amplification, inducing a substantially larger $E_z$ in the channel for the same $V_G$.

The L-K equation, which describes the polarization ($P$) and electric field in a ferroelectric material ($E_{FE}$), is given by

$$E_{FE} = 2\alpha_{FE}P + 4\beta_{FE}P^3. \tag{1}$$

Where $\alpha_{FE}$ and $\beta_{FE}$ are the ferroelectric material parameters. The device operates under a voltage division condition, $V_G = V_{FE}(E_Z) + \psi_s$, where $V_{FE}$ and $\psi_s$ are the voltages across the ferroelectric layer and the surface potential of TI channel, respectively. The detail derivation of $V_{FE}$ is documented in the Supporting Information 4. Figure 2c shows the $E_z$ as a function of $V_G$ for effects by various gate insulators. Notably, this relationship of $V_G$ makes maximum electric field amplification condition, when ferroelectric material thickness $t_{FE} = 1/(2\alpha_{FE}C_{ch})$, where $C_{ch}$ is channel capacitance. Within the optimized $t_{FE}$, maximum gate voltage condition ($V_{G,max}$) is calculated as

$$V_{G,max} = \frac{C_{ch}^2 t_{ch}^3}{P_r^2} E_z^3. \tag{2}$$

Where $t_{ch}$ is channel thickness, and $P_r$ is remanent polarization of the ferroelectric material, which $P_r$ is equal to $P_{r,max}$ in NC-TIFETs condition.

This equation provides a powerful design principle: a higher $P_r$ dramatically reduces the gate voltage required to switch the device. As predicted by Eq. (2), bismuth iron oxide[25,29] (BiFeO₃, BFO), with its exceptionally high $P_r$, is capable of reaching the $E_c$ at an ultra-low $V_G$. However, BFO presents a critical manufacturing challenge: its NC performance is acutely sensitive to the $t_{FE}$ and deviating from the optimal thickness leads to a loss of amplification or undesirable hysteresis (see Figure S3 of the Supporting Information). Therefore, we selected HZO[28] as the gate insulator. HZO is a hafnium-based dielectric employed in contemporary semiconductor devices, ensuring its compatibility with existing manufacturing processes[30]. Crucially, HZO is compatible with atomic layer deposition (ALD), which represents a superior engineering conformality, large-area uniformity, and atomic level thickness control[31]. We expect that utilizing of ALD for the fabrication of HZO leads to optimized $t_{FE}$, thereby achieving stable, hysteresis-free NC operation.

Our HZO ferroelectric parameters used in this study are based on room-temperature values[28]. At cryogenic temperatures, however, the coercive field ($E_{co}$) typically increases, which can shift the optimized $t_{FE}$ and the hysteresis window. To address this, we systematically investigate the impact of $E_{co}$ variations on the $t_{FE}$ optimization and the hysteresis window in the Supporting Information 7 (Figure S3 and S4). Furthermore, the impact of $P_r$ on switching characteristics is demonstrated in Figure S6 of Supporting Information.

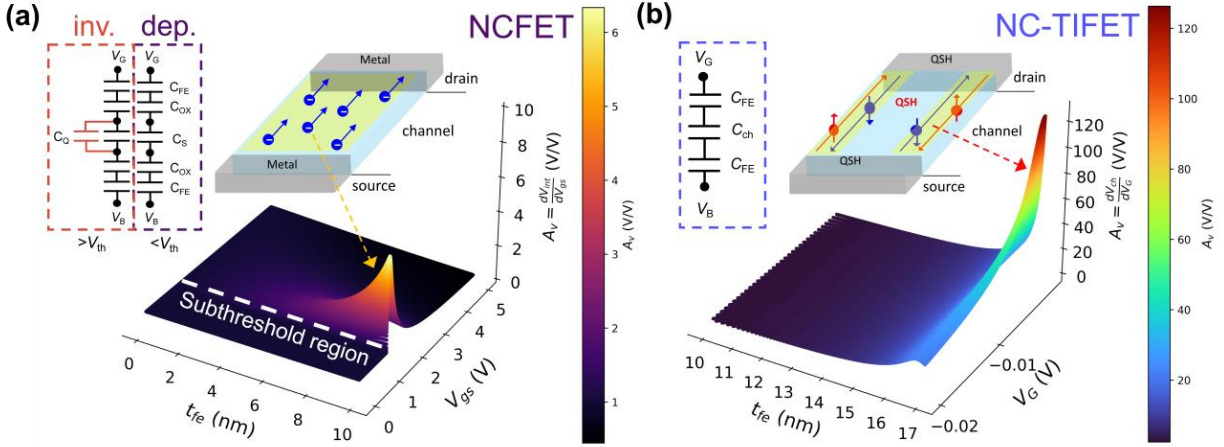

**Figure 3.** Comparison of voltage gains $A_V$ between an NCFET and an NC-TIFET. (a) The voltage gain ($A_V = dV_{int}/dV_{GS}$) of the NCFET with a Si channel and an HZO ferroelectric layer, plotted as a function of $t_{FE}$ and gate-source voltage ($V_{GS}$). The subthreshold region is indicated by the white dashed line. (b) The voltage gain ($A_V = dV_{ch}/dV_G$) of the NC-TIFET with a 1T′-MoS$_2$ channel and an HZO ferroelectric layer, is shown as a function of $t_{FE}$ and $V_G$. Insets for each device depict the equivalent capacitance circuit (top-left) and a schematic of device operation (center). 2D cross-sectional plots Gain vs $V_G$ are provided in Figure S5 of Supporting Information.

NCFETs have been considered promising candidates for future low-power applications due to their ability to achieve sub-60 mV/dec SS, which is attributed to the polarization response of the ferroelectric material that amplifies the gate voltage[30,32,33]. However, NCFETs are known that have critical hysteresis issue, if the capacitance matching process is not correctly performed[17]. The main element of hysteresis is the $C_Q$, which increases rapidly after bulk charge inversion, mainly occur in modern FETs with ultra-thin channels. (see equivalent capacitance circuit in Figure 3a). Therefore, even if capacitance matching is achieved to minimize hysteresis, the $A_V$ induced by NC is significantly limited in the sub-threshold region due to the limited availability of charges in the gate stack to promote voltage growth (see Figure 3a). In contrast, NC-TIFETs modulate only the edge states during switching, thereby avoiding charge generation within the bulk. In other words, unlike NCFETs, NC-TIFETs exhibit a channel behavior similar to that of a dielectric material, meaning that the channel capacitance is largely independent of the gate bias. This characteristic leads to a series capacitance circuit that is primarily influenced by the ferroelectric capacitance and the channel capacitance (see equivalent capacitance circuit in Figure 3b). As a result, NC-TIFETs can achieve hysteresis-free operation under optimal $t_{FE}$ conditions, which allows for the realization of maximum $A_V$ values.

While TIFETs avoid the $C_Q$ bottleneck via the absence of bulk inversion, the topological edge states introduce a finite $C_Q$ at the edges. Crucially, these edge states are robust features protected by the bulk band topology. As a result, the edge $C_Q$ is expected to be negligible with respect to the inversion-symmetry breaking in the bulk, while the topologically protected edges ensure stable transport. To accurately model the electrostatics including edge states, it is essential to employ rigorous two-dimensional electrostatic simulations that account for the parallel geometric contributions of the conducting edges and the insulating bulk under vertical electric gate fields. Consequently, we identify the precise quantification of edge $C_Q$ and its impact on capacitance scaling as a subject for future investigation.

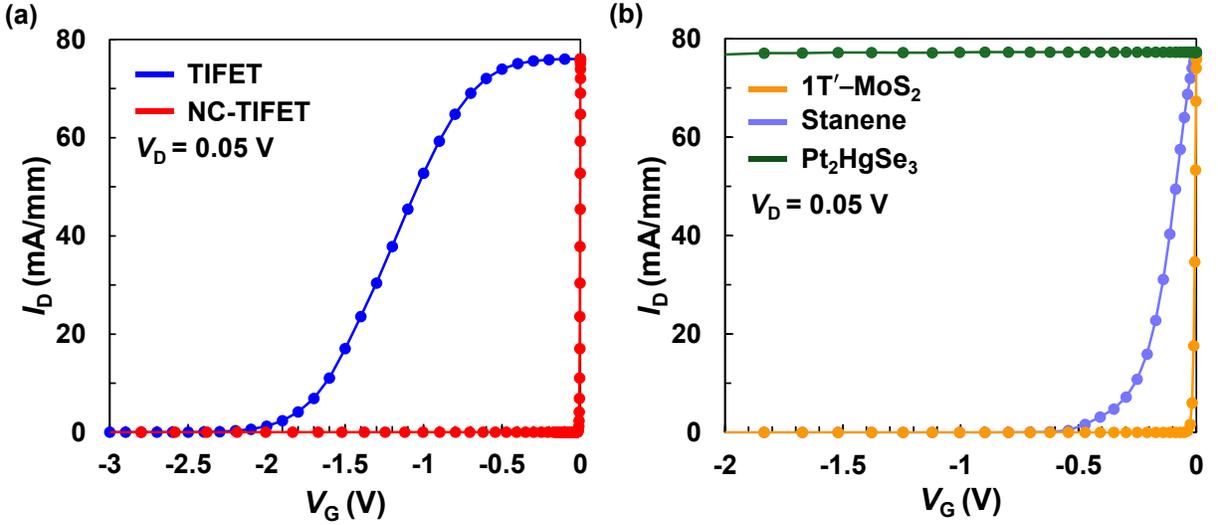

**Figure 4.** Transfer characteristics and the effect of various topological channel materials. (a) Comparison of transfer characteristics for a conventional TIFET (blue) and an NC-TIFET (red) at $V_D = 0.05\ V$. Both devices utilize a 1T′-MoS$_2$ channel. (b) Transfer characteristics of NC-TIFETs using three different topological channel materials: 1T′-MoS$_2$, stanene, and Pt$_2$HgSe$_3$. The ferroelectric gate is HZO for all devices.

Due to the superior $A_V$ in NC-TIFETs, NC-TIFETs can result in an extremely steep-slope transfer curve. TIFETs require a large switching voltage to induce a topological phase transition, because of positive voltage drop of dielectric material when induced gate bias (see Figure 2a). In contrast, the negative voltage drop in NC layer and hysteresis-free operation of the NC-TIFET allows to a dramatically steepened transfer curve (Figure 4a), and powerful output characteristics (see Figure S7 of supporting information). Next, we investigated the role of the channel material on device performance by comparing three different 2D topological insulators: 1T′-MoS$_2$[20,21], stanene[19], and Pt$_2$HgSe$_3$[34,35] (Figure 4b). Stanene has been identified as a promising 2D TI material due to its small band gap and a large material-specific parameter that influences the $E_c$[19]. However, NC-TIFETs based on stanene channel material result in a switching voltage of approximately 0.5 V, offering no significant power dissipation advantage over conventional HEMTs. $Pt_2HgSe_3$ has also been identified as a promising 2D TI material, because of large band gap to prevent thermal leakage. However, $Pt_2HgSe_3$ of large band gap (160 meV[34]) and relatively low electric field potential parameter (i.e., $\alpha_V$ and $\alpha_R$, see Table S2 of supporting information.) impedes the device from switching OFF within extremely low voltage range. In contrast, due to small band gap and high material specific parameter effecting for $E_z$[21] (see Table S1 of supporting information.), 1T′-$MoS_2$ get ultra-low switching voltage, (see Figure 5) which emerges as a strong candidate for cryogenic applications. The model parameters for these topological materials are provided in Figure S2 of the Supporting Information.

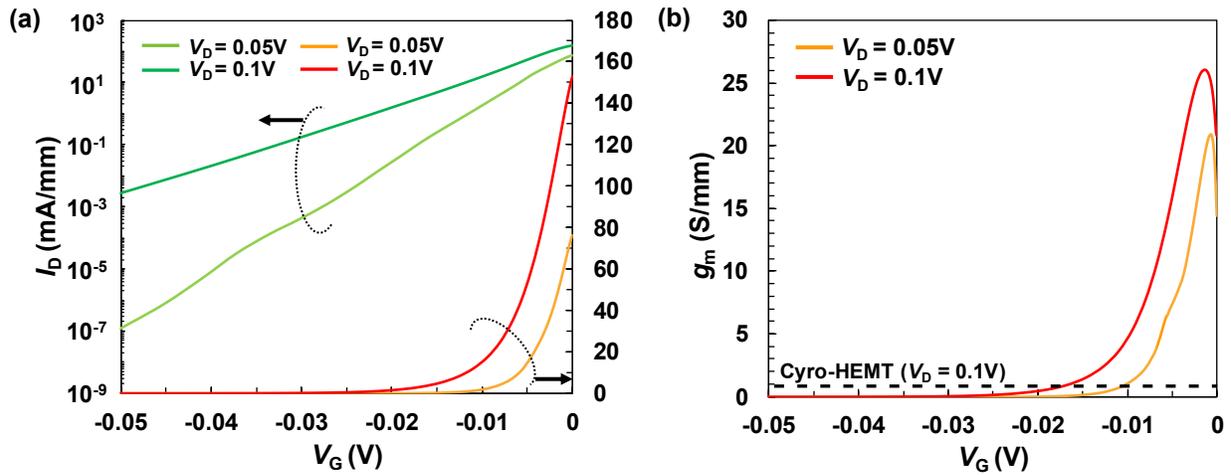

**Figure 5.** Electrical characteristics of NC-TIFETs. (a) Transfer characteristics ($I_D$–$V_G$) of the TIFET and NC-TIFET at $V_D = 0.05\ V\ and\ 0.1\ V$. (b) Transconductance characteristics ($g_m$ – $V_G$) at $V_D = 0.05\ V\ and\ 0.1\ V$, benchmarked against experimental data from a Cryo-HEMT[8]. All of data were obtained at $T$ = 4K. For all results shown in Figure 5, the channel and ferroelectric materials are 1T′-$MoS_2$ and HZO, respectively.

As the result, the optimized combination of 1T′-MoS$_2$ and HZO Combination results in an extremely steep-slope transfer curve. As shown in Figure 5a, the NC-TIFET achieves a sub-20 mV switching voltage (defined for an ON/OFF ratio of 10$^3$) even at a low drain bias of $V_D = 0.05$ V. Notably, despite this low operating $V_D$, rapid switching capability of NC-TIFETs produces an exceptionally high $g_m$ of approximately 26 S/mm at $V_D = 0.1$ V. This value is over an order of magnitude higher than the experimentally recorded $g_m$ of a state-of-the-art Cryo-HEMT (0.8 S/mm)[8] (Figure 5b). It is important to acknowledge that the experimental Cryo-HEMT data reflect parasitic effects and non-idealities—such as interface traps and gate leakage—which inevitably degrade $g_m$, whereas our model captures the intrinsic performance limits. However, even after accounting for these extrinsic factors, the combination of an ultra-low switching voltage and high transconductance establishes the NC-TIFET as a compelling candidate for cryogenic electronic interfaces, including low-noise amplifiers (LNAs) and analog-to-digital converters (ADCs).

While the presented NC-TIFET, modeled with DFT-derived parameters, demonstrates exceptional performance, it is important to note that this represents an idealized case. Therefore, in order for the development of such technology to proceed, it is necessary that several fundamental issues be identified and addressed at the structural, device, and contact levels. First, at the structural level, stack engineering should be explored. While our single-layer ferroelectric design demonstrates excellent switching, the integration of multi-layer stacks has emerged as a promising strategy to enhance ON-currents and decreased gate capacitance. Second, to validate the device's potential for analog applications such as amplifiers, a thorough analysis of its high-frequency (RF) and noise characteristics is imperative. To this end, it is imperative to incorporate the effects of parasitic capacitance within the TIFET, necessitating the precise calculation of the transit frequency ($f_T$), power consumption ($P_{DC}$), and noise indication factor is important. Especially, conventional noise indication factor for charge-based devices such as HEMTs (i.e., the $\sqrt{I_D}g_m^{-1}$ factor[8,36]) may not be directly applicable due to the characteristics of spin current based topological edge states. Therefore, a novel noise model may be necessary, incorporating unique spin-based transport and realistic impurity scattering mechanisms, including magnetic impurities at the edge states[37,38]. Third, for practical circuit applications such as LNAs or DACs/ADCs, realizing complementary p-type characteristics is essential. Although realistic modeling of p-type TIFETs remains unexplored, we suggest that p-type operation could be theoretically achieved by inverting the double-gate bias configuration (i.e., swapping the roles of the top and bottom gate voltages). This inversion enables p-type modulation without physical doping. However, further research is required to fully elucidate the operational mechanisms of p-type TIFETs and develop specific integration schemes for CMOS architectures. Finally, the practical implementation of the source/drain (S/D) contacts warrants discussion. While our theoretical model assumes ideal contacts where the S/D regions are integrated with the same topological channel, the physical interface is critical for effectively exploiting topological edge states. We propose two viable approaches: (1) phase engineering, utilizing 1T-1T′ heterostructures as demonstrated in recent 1T-2H heterophase studies[39,41], where the metallic phase 1T can effectively decrease $R_C$ and enhance device current[40,41]. (2) One-dimensional (1D) edge contacts, which have been proposed as a key solution for reducing the contact resistance[42,43]. Furthermore, our analytical model incorporating finite $R_C$ (Supporting Information 10) confirms that the NC gain remains robust, preserving steep switching characteristics even under realistic contact conditions (Figure S8). Therefore, overcoming these multilevel will be crucial for realizing the full potential of NC-TIFETs as high-performance cryogenic devices for large-scale quantum computing systems. In conclusion, we have theoretically demonstrated that the NC-TIFET architecture overcomes the fundamental quantum capacitance bottleneck of NCFETs, enabling superior switching performance by fully exploiting the NC effect. As a result, our device achieves superior switching characteristics, including a sub-20 mV switching voltage (at $V_D = 0.05\ V$) and an ultra-high transconductance of 26 S/mm (at $V_D = 0.1$ V), establishing the NC-TIFET as a key technology for the cryogenic electronic interfaces for large-scale quantum computing, such as LNAs and ADCs.–However, while gate-tunable 1T'-MoS$_2$ has been experimentally observed[44], translating these theoretical advantages into reality requires addressing the metastability of the 1T'-MoS$_2$ phase[45,46,47,48]. Future efforts must focus on robust fabrication strategies—such as surface passivation[49,50], control of magnetic impurities in edge states[37,38], and defect engineering[50]—to stabilize the 1T' phase against NC-induced high internal fields and ensure reliable operation. Consequently, establishing a stable fabrication process for 1T'-MoS$_2$ remains the critical milestone for realizing high-performance NC-TIFETs in large-scale cryogenic quantum computing systems.

**ASSOCIATED CONTENT**

Supporting Information

The Supporting Information is available free of charge at

https://pubs.acs.org/doi/10.1021/acs.nanolett.5c06182.

Device parameters, model method in NC-TIFETs, DFT parameters of 1T'-MoS$_2$, Pt$_2$HgSe$_3$, and stanene, L–K equation used to model NC-TIFETs, optimization of the NC-TIFET channel length and width, model parameter fitting for DFT, optimization of the ferroelectric thickness ($t_{FE}$) in NC-TIFETs in cryogenic environments, effects of different ferroelectric materials in NC-TIFETs, output characteristics of NC-TIFETs, and formulas for the TIFET drain current equation combined with contact resistance (PDF)

## AUTHOR INFORMATION

Corresponding Author

**Yeonghun Lee -** Department of Intelligent Semiconductor Engineering, Incheon National University, 119 Academy-ro, Yeonsu-gu, Incheon 22012, Republic of Korea; Department of Electronics Engineering, Incheon National University, Incheon National University, 119 Academy-ro, Yeonsu-gu, Incheon 22012, Republic of Korea; Research Institute for Engineering and Technology, Incheon National University, 119 Academy-ro, Yeonsu-gu, Incheon 22012, Republic of Korea; orcid.org/0000-0002-6058-1316; Email: y.lee@inu.ac.kr

Authors

**Yosep Park -** Department of Intelligent Semiconductor Engineering, Incheon National University, 119 Academy-ro, Yeonsu-gu, Incheon 22012, Republic of Korea; orcid.org/0009-0003-0123-8942

**Yungyeong Park -** Department of Electronics Engineering, Incheon National University, Incheon National University, 119 Academy-ro, Yeonsu-gu, Incheon 22012, Republic of Korea; orcid.org/0009-0008-6621-4796

**Hyeonseok Choi -** Department of Electronics Engineering, Incheon National University, Incheon National University, 119 Academy-ro, Yeonsu-gu, Incheon 22012, Republic of Korea; orcid.org/0009-0006-1523-5605

**Subeen Lim -** Department of Electronics Engineering, Incheon National University, Incheon National University, 119 Academy-ro, Yeonsu-gu, Incheon 22012, Republic of Korea; orcid.org/0009-0002-1957-2677

Complete contact information is available at:

https://pubs.acs.org/10.1021/acs.nanolett.5c06182

Author Contributions

Y.L. supervised the projects. Yo.P., Yu.P., H.C., and S.L. performed the theoretical calculation and data analysis. Yo.P., Yu.P., and Y.L. wrote the manuscript. All authors have discussed the results, read the manuscript, and agreed with its content.

Notes

The authors declare the following competing financial interest(s): Yo.P., Yu.P., and Y.L. hold two Korean patents (Field Effect Transistor Using Phase Transition Channel Material; and Apparatus and System for Quantum Computing Including Quantum Transistor) related to the technology described in this work.

**ACKNOWLEDGMENT**

This work was supported by the National Research Foundation of Korea (NRF) grant funded by the Korea government (MSIT) (No. 2022R1F1A1073068).

Supporting Information for

"Designing Extremely Low-Power Topological Transistors with 1T′–MoS$_2$ and HZO for Cryogenic Applications"

Yosep Park[1], Yungyeong Park[2], Hyeonseok Choi[2], Subeen Lim[2], and Yeonghun Lee[1,2,3]

[1]Department of Intelligent Semiconductor Engineering, Incheon National University, Incheon 22012, Republic of Korea.

[2]Department of Electronics Engineering, Incheon National University, Incheon 22012, Republic of Korea.

[3]Research Institute for Engineering and Technology, Incheon National University, Incheon 22012, Republic of Korea.

e-mail: y.lee@inu.ac.kr

**Contents**

**Supporting Information 1. Device parameters**

Section 1-1) TIFET parameters

Section 1-2) Ferroelectric material parameter in NC-TIFETs

**Supporting Information 2. Model Method in NC-TIFETs**

Section 2-1) Hamiltonian-based continuum ($\mathbf{k} \cdot \mathbf{p}$) model (1T′-MoS$_2$)

Section 2-2) Hamiltonian-based Kane–Mele model (Pt$_2$HgSe$_3$, Stanene)

**Supporting Information 3. DFT parameters of 1T′-MoS$_2$, Pt$_2$HgSe$_3$, and stanene**

**Supporting Information 4. Landau-Khalatnikov (L–K) equation to model NC-TIFETs**

**Supporting Information 5. Optimization of NC-TIFET channel length and width**

**Supporting Information 6. Model parameter fitting for DFT**

**Supporting Information 7. Optimization of ferroelectric thickness ($t_{\mathrm{FE}}$) in NC-TIFETs in cryogenic environments**

**Supporting Information 8. Effects of different ferroelectric materials in NC-TIFETs.**

**Supporting Information 9. Output characteristics of NC-TIFETs**

**Supporting Information 10. Formulas for TIFET drain current equation combined with contact resistance.**

## Supporting Information 1. Device parameters

The TIFET (NC-TIFET) model is based on the work of Park et al[1]. The device parameters presented in Sections 1-1 and 1-2 were directly extracted from our device structure.

## Section 1-1) TIFETs parameters

| symbol | Parameter | Value |
|--------|-----------|-------|
| $E_C$ | Critical electric field point | 1.3 V/Å |
| $E_g$ | Channel material bandgap | 43 meV |
| $\delta_p$ | Lowest conduction band structure energy | - 0.445 eV |
| $\delta_d$ | Highest conduction band structure energy[2] | - 0.132 eV |
| $m_x^d/m_0$ | Effective mass valence band in x-direction[2] | 0.92 |
| $m_x^p/m_0$ | Effective mass conduction band in x-direction[2] | 0.29 |
| $m_y^d/m_0$ | Effective mass valence band in y-direction[2] | 2.32 |
| $m_y^p/m_0$ | Effective mass conduction band in y-direction[2] | 0.48 |
| $v_1$ | Velocity x direction[2] | $0.23 \times 10^5$ m/s |
| $v_2$ | Velocity y direction[2] | $3.383 \times 10^5$ m/s |
| $\alpha$ | material specific electrical parameter[2] | 0.21 eÅ |
| $t_{ch}$ | 1T′–MoS$_2$ thickness | 0.5063 nm |
| EOT | Equivalent oxide thickness | 0.5 nm |
| $\varepsilon_r$ | The relative permittivity of 1T′-MoS$_2$ | 2.4547 |
| $\varepsilon_{ox}$ | The relative permittivity of oxide | 3.9 |

Table S1. Summary of calculated TIFET parameters with a 1T'-MoS$_2$ channel. The dielectric material is SiO$_2$. $m_0$ is the rest mass of electron.

| symbol | Parameter | Value |
|---|---|---|
| $E_C$ | Critical electric field point | 0.7 (0.93) V/Å |
| $E_g$ | Channel material bandgap | 62.1 (143.3) meV |
| $a$ | Lattice constant | 0.467 (0.75) nm |
| $t$ | Hopping | 0.7 (0.2) eV |
| $\lambda_{SO}$ | Spin-orbit coupling parameter | 35/(3$\sqrt{3}$) (77.5/(3$\sqrt{3}$)) meV |
| $\alpha_V$ | Staggered potential parameter | $5 \times 10^{-12}$ ($7 \times 10^{-12}$) eV |
| $\alpha_R$ | Rashba parameter | 0 ($2 \times 10^{-12}$) eV |
| $t_{ch}$ | Stanene thickness | 0.3229 (0.5191) nm |
| EOT | Equivalent oxide thickness | 0.5 nm |
| $\varepsilon_r$ | The relative permittivity of channel | 1.324 (6.048) |
| $\varepsilon_{ox}$ | The relative permittivity of oxide | 3.9 |

Table S2. Summary of calculated TIFETs channel material parameters for stanene. Parameters for Pt$_2$HgSe$_3$ are shown in parentheses. The dielectric material is SiO$_2$.

## Section 1-2) Ferroelectric material parameter in NC-TIFET table

| symbol | Parameter | Value |
|---|---|---|
| $\alpha_{FE}$ | Ferroelectric material parameters[3] | $-6.8 \times 10^8$ mF$^{-1}$ |
| $\beta_{\text{FE}}$ | | $2.87 \times 10^9 \text{ m}^5\text{F}^{-1}\text{C}^{-2}$ |
| $T_{\text{FE\_OPT}}$ | Optimized Ferroelectric thickness | 17.135 nm |
| $P_r$ | Remanent polarization[3] | 34.4 µC/cm$^2$ |
| $E_{\text{C0}}$ | Coercive field[3] | 1.8 MV/cm |

Table S3. Summary of calculated ferroelectric material parameters for HfZrO$_2$ (HZO) in NC-TIFETs. The optimized ferroelectric thickness which maximizes voltage amplification, $T_{\text{FE\_OPT}} = -1/2\alpha_{\text{FE}}C_{\text{ch}}$, where $\alpha_{\text{FE}}$ is the ferroelectric material parameter and $C_{\text{ch}}$ is the channel capacitance (using the 1T′-MoS$_2$ capacitance value). $T_{\text{FE\_OPT}}$ is included both top and bottom layer. Note that the ferroelectric parameters ( $P_r$ , $E_{\text{C0}}$ ) listed above were extracted from P-E hysteresis measurements conducted at a frequency of 1 kHz.

**Supporting Information 2. Model Method in NC-TIFETs**

**Section 2-1) Hamiltonian-based continuum (k · p) model (1T′-MoS₂)**

To investigate the characteristics of the TIFETs with 1T′-MoS$_2$ channel, We employed a continuum **k · p** Hamiltonian and parameters derived from Das et al[2]. In this study, the parameters α and $\delta_p$ were updated to reproduce the field-induced topological phase transition with precise consideration of charge redistribution under the electric field applied along the $z$ direction ($E_z$). The **k · p** model were implemented using the Kwant Python package[4].

**Section 2-2) Hamiltonian-based Kane–Mele model (Pt₂HgSe₃, Stanene)**

We implemented the Kane-Mele tight-binding model using the Kwant Python package[4]. For details regarding the Hamiltonian and parameters for Pt$_2$HgSe$_3$ and stanene, please refer to Park et al[1]. Although Pt$_2$HgSe$_3$ was not calculated in the study by Park et al., the parameter types and extraction method used in our study are consistent with those described in their work[1].

**Supporting Information 3. DFT parameters of 1T′-MoS₂, Pt₂HgSe₃, and stanene**

All channel materials used in the TIFET model, First-principles calculations based on DFT[5,6] were conducted using the pseudo-atomic orbital (PAO) based software package OpenMX[7,8] with 220 Ry of real-space grid cutoff energy and the plane wave (PW) based software package VASP[9–13] with 500 eV of plane-wave cutoff energy. For both codes, the exchange-correlation functional was described within the generalized gradient approximation (GGA) as formulated by Perdew, Burke, and Ernzerhof (PBE)[14]. The self-consistent field (SCF) loop convergence criterion was set to $10^{-6}$ eV for both codes. $9{\times}9{\times}1$ and $9{\times}1{\times}1$ meshes were used for bulk and nanoribbon calculations, respectively, to sample the first Brillouin zone. Non-collinear spin texture and SOC effects were considered in all calculations. To eliminate spurious interactions between monolayers, a vacuum region exceeding 15 Å was added along the z-direction for both bulk and nanoribbon calculations. For atomic position relaxation, the force tolerance was set to 0.015 eV/Å for OpenMX and 0.01 eV/Å for VASP. The atomic positions and cell volume of the bulk unit cell were optimized while keeping the cell shape fixed.

**Supporting Information 4. Landau-Khalatnikov (L–K) equation to model NC-TIFETs**

The voltage drop across the ferroelectric layer in the NC-TIFET structure is modeled using the Landau−Khalatnikov (L-K) equation[15], following the approach proposed in ref[16–18],

$$V_{\text{FE}} = t_{\text{FE}}E_{\text{FE}} = 2\alpha_{\text{FE}}t_{\text{FE}}P + 4\beta_{\text{FE}}t_{\text{FE}}P^3, \tag{4}$$

where $V_{\text{FE}}$ is the voltage drop across the ferroelectric layer, $E_{\text{FE}}$ is the electric field within the ferroelectric material, $P$ is polarization of the ferroelectric material, $\alpha_{\text{FE}}$ and $\beta_{\text{FE}}$ are the ferroelectric material parameters, and $t_{\text{FE}}$ is Ferroelectric material thickness (included both top and bottom layer). The values of $\alpha_{\text{FE}}, \beta_{\text{FE}}$ and $t_{\text{FE}}$ are given in Supporting Information Table 3.

The polarization, $P$, can be approximated as the charge $Q \approx P$, where $Q = \psi_s C_{ch}$, $\psi_s = E_{ch}t_{ch}$ is surface potential of channel, $C_{ch}$ is channel capacitance, $E_{ch}$ is channel electric field ,and $t_{\text{ch}}$ is channel thickness.

From series connection of the Ferroelectric layer and topological insulator (TI) (see the small-signal model of the NC-TIFET in Figure 3c of the main text), the gate voltage is described by

$$V_{\text{G}} = V_{\text{FE}} + \psi_s = (t_{\text{ch}} + 2\alpha_{\text{FE}}t_{\text{FE}}t_{\text{ch}}C_{\text{ch}})E_{\text{ch}} + 4\beta_{\text{FE}}t_{\text{FE}}(C_{\text{ch}}t_{\text{ch}}E_{\text{ch}})^3, \tag{5}$$

Differentiating Equation (5) with respect to $E_{\text{ch}}$ at $V_{\text{G}}$ (i.e., $dV_{\text{G}}/dE_{\text{ch}}$ at $E_{\text{ch}} = 0$), we can obtain the optimum $t_{\text{FE}} = 1/(2\alpha_{\text{FE}}C_{\text{ch}})$, which reaches the maximum electric field condition. To satisfy $t_{\text{FE}}$, this maximum gate voltage condition ($V_{\text{G,max}}$) is calculated as

$$V_{\text{G,max}} = \frac{C_{\text{ch}}^2 t_{\text{ch}}^3}{P_{\text{r}}^2}E_z^3. \tag{6}$$

where $P_{\text{r}}$ is remanent polarization of the ferroelectric material, which $P_{\text{r}}$ is equal to $P_{\text{r,max}}$ in NC-TIFETs condition.

Also, from the equation (4), the $P_r$ can be derived when the $E_{\mathrm{FE}}$ is zero, as expressed by

$$E_{\mathrm{FE}} = 2\alpha_{\mathrm{FE}}P + 4\beta_{\mathrm{FE}}P^3 = 0, \tag{7}$$

Upon implementation as an expression for P, the following $P_r$ is can be obtained.

$$P_r = \sqrt{-\frac{\alpha_{\mathrm{FE}}}{2\beta_{\mathrm{FE}}}}. \tag{8}$$

**Supporting Information 5. Optimization of NC-TIFET channel length and width**

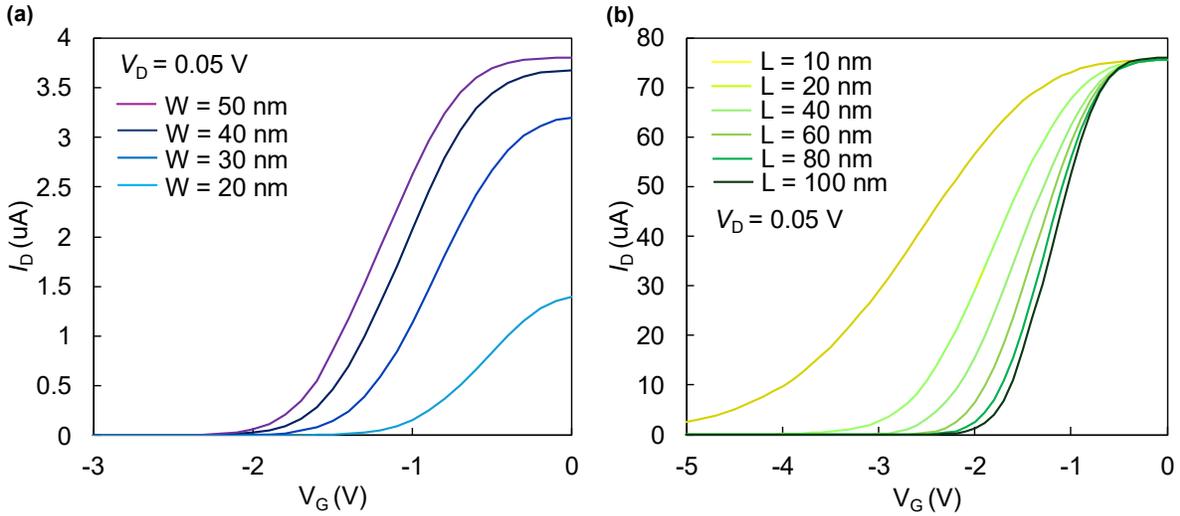

**Figure S1.** Effects of channel width and channel length modulation in TIFETs.

(a) Transfer characteristics as a function of channel width, ranging from 20 to 50 nm fixed at $V_D = 0.05$V. (b) Transfer characteristics as a function of channel length, ranging from 10 to 100 nm. In Figure S1, insufficient channel width can disrupt the unique topological edge states due to the finite-size effect[19,20]. However, this effect is suppressed as the width increases, and the device characteristics converge at a width of 50 nm, allowing the topological edge states. Additionally, an excessively short channel can induce significant tunneling leakage[1]. Therefore, $W = 50$ nm and $L = 100$ nm is best transfer characteristics for the TIFETs.

.

**Supporting Information 6. Model parameter fitting for DFT**

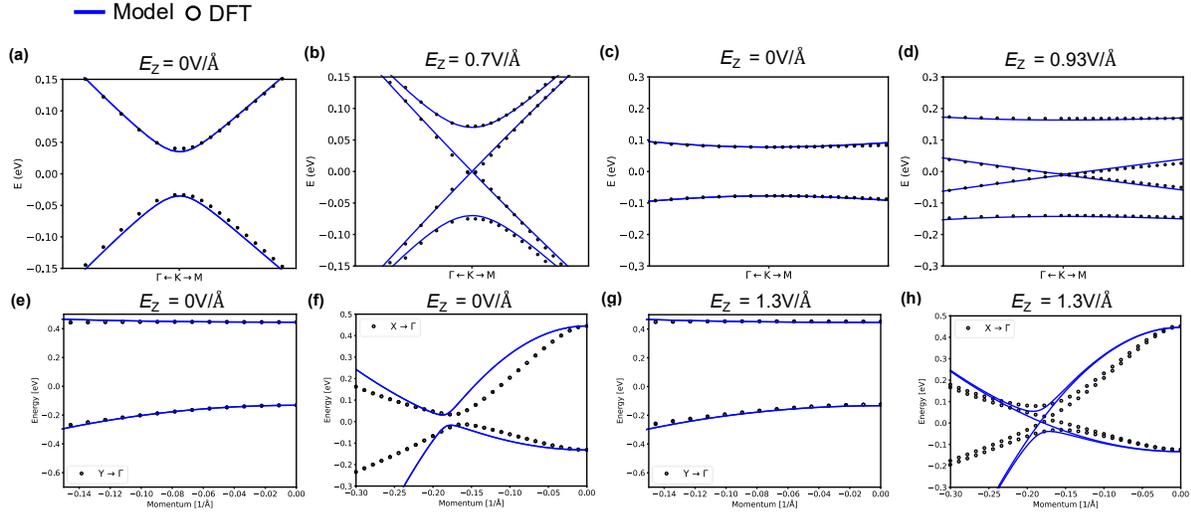

**Figure S2.** Parameter extraction for various topological channel materials.

(a-d) The bulk band structures of (a,b) stanene, and (c,d) $Pt_2HgSe_3$ have been calculated using a tight-binding (TB) model. The TB model (solid lines) is fitted to density functional theory (DFT) calculations (open circles) around K-point under different critical electric fields. (e-h) Bulk band structures of $1T'$-$MoS_2$, calculated using a $\mathbf{k} \cdot \mathbf{p}$ Hamiltonian. The $\mathbf{k} \cdot \mathbf{p}$ model (solid lines) is fitted to DFT calculations (open circles) along the X–$\Gamma$ and Y–$\Gamma$ high-symmetry paths. The model parameters were extracted at the ON-state ($E_z$ = 0 V/Å) and their respective critical points ($E_z$ = 0.7 V/Å, 0.93 V/Å, and 1.3 V/Å).

**Supporting Information 7. Optimization of Ferroelectric thickness ($t_{FE}$) in NC-TIFET in cryogenic environments**

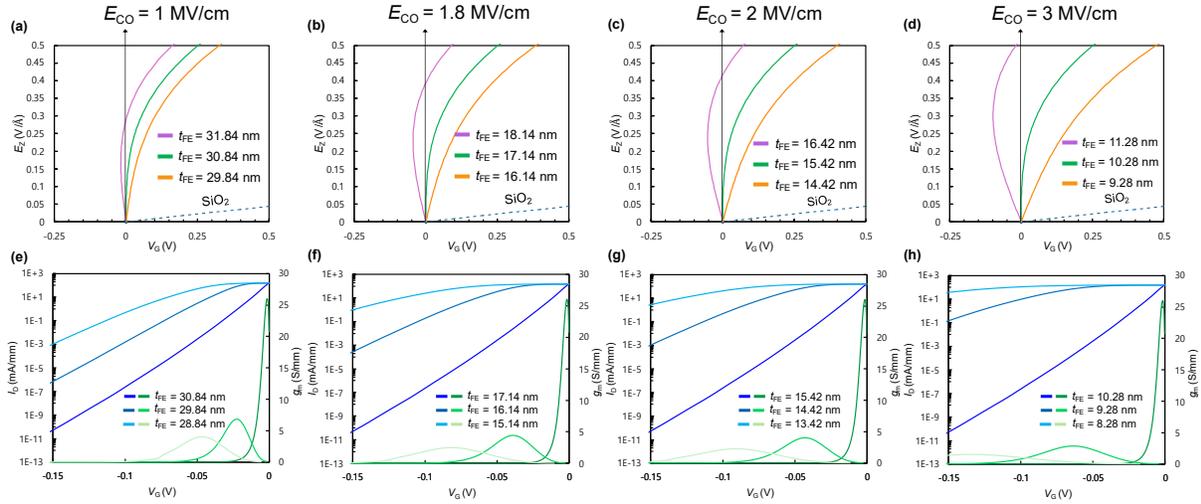

**Figure S3.** Effects of $t_{FE}$ variation in NC-TIFETs under different coercive field ($E_{co}$) conditions. (a-d) $E_Z$ as a function of $V_G$ with different $t_{FE}$ values for $E_{co}$ = 1, 1.8, 2, and 3 MV/cm, respectively. The dashed line represents the $E_Z$ for a TIFET with $SiO_2$ gate dielectric with a thickness of 0.5 nm. (e-h) Corresponding device characteristics as a function of $t_{FE}$ at $V_D$ = 0.1 V. The left side is log-scale transfer characteristics, right side shows the linear-scale of transconductance characteristics.

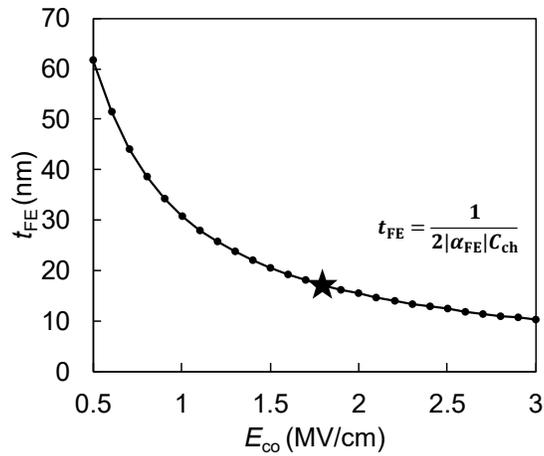

**Figure S4.** Optimized ferroelectric thickness ($t_{\text{FE\_OPT}}$) as a function of coercive field ($E_{\text{co}}$). The remanent polarization ($P_r$) is fixed at the HZO value of 34.4 $\mu\text{C}/\text{cm}^2$. The star symbol indicates the parameter used for our NC-TIFET simulation ($E_{\text{co}} = 1.8$ MV/cm).

Optimized ferroelectric thickness $t_{\text{FE\_OPT}} = 1/2\alpha_{\text{FE}}C_{\text{ch}}$ in NC-TIFETs is critical because it governs the amplification of electric field within the ferroelectric layer. As shown in Figure S3(a-d), if the thickness exceeds the optimal range, it can lead to hysteresis. Conversely, if the thickness falls short of the optimal range, it cannot achieve sufficient negative capacitance (NC) effects. As shown in Figure S3(e-h), These electric field amplification characteristics directly impact the steepness of the switching characteristics in NC-TIFETs. Also, these figure illustrates how the device characteristics ($I_{\text{D}} - V_{\text{G}}$ and $g_m$) evolve as $E_c$ varies, highlighting that a thinner ferroelectric layer is required at cryogenic temperatures to maintain the non-hysteretic steep switching performance.

In cryogenic environments, the coercive field ($E_{\text{co}}$) of the ferroelectric layer is known to increase compared to room temperature[21,22]. As shown in Figure S3 and S4, an increase in $E_{\text{co}}$ (with fixed HZO $P_r$) leads to a larger magnitude of the ferroelectric coefficient $|\alpha_{FE}|$, since $|\alpha_{FE}| \propto E_{\text{co}}/P_r$. Consequently, the optimal thickness ($t_{\text{FE\_OPT}} = 1/(2|\alpha_{FE}|C_{ch})$) required to maximize voltage gain without hysteresis decreases as the temperature drops. Consequently, the device performance becomes more sensitive to variations in ferroelectric thickness as $E_{\text{co}}$ increases.

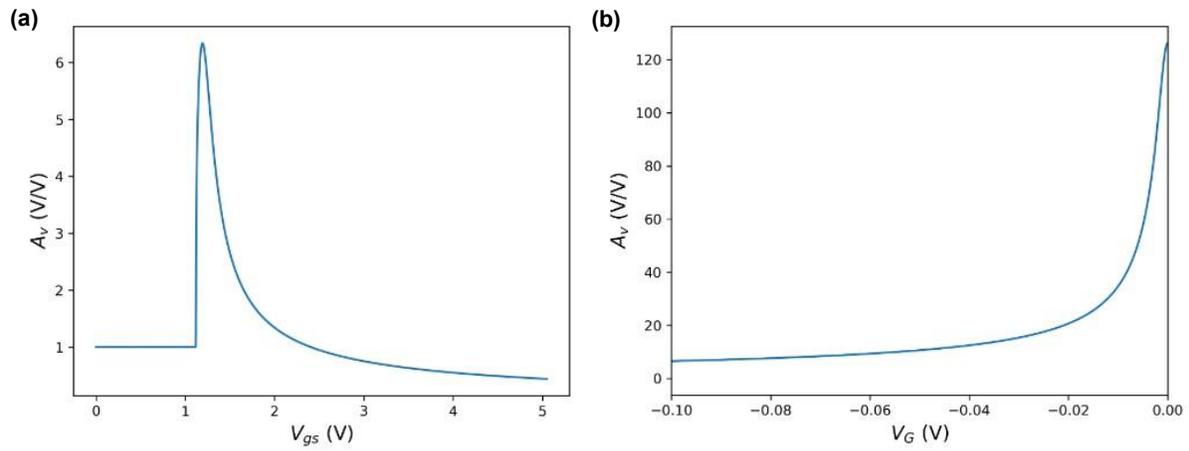

**Figure S5.** 2D cross-sectional plots ($A_V$ vs $V_G$) of the manuscript of Figure 3. (a) Left figure is NCFET, and (b) right figure is NC-TIFET.

**Supporting Information 8. Difference of ferroelectric material in NC-TIFETs.**

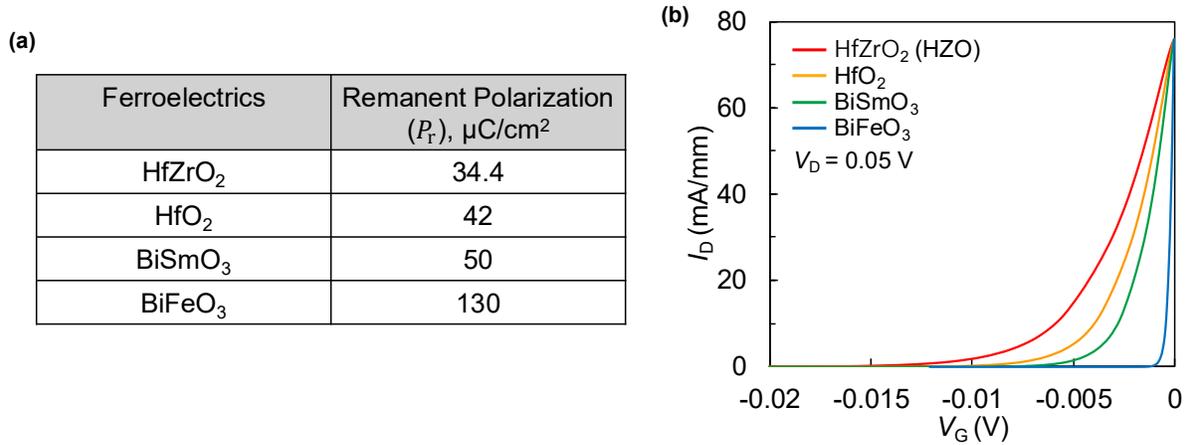

(a)

| Ferroelectrics | Remanent Polarization ($P_r$), μC/cm² |
|---|---|
| HfZrO₂ | 34.4 |
| HfO₂ | 42 |
| BiSmO₃ | 50 |
| BiFeO₃ | 130 |

**Figure S6.** Impact of Remanent Polarization on Switching Characteristics. (a) Comparison of remanent polarization ($P_r$) for various ferroelectric materials: $HfZrO_2$ (HZO)[3], $HfO_2$[23], $BiSmO_3$[24], and $BiFeO_3$ (BFO)[25]. (b) Transfer characteristics of NC-TIFETs utilizing different ferroelectrics materials. For all simulations, 1T′-MoS₂ was used as the channel material. Note that all of the ferroelectric parameters are at room temperature.

A larger $P_r$ leads to higher amplification, resulting in a steeper slope of transfer characteristics. Among the materials investigated, $BiFeO_3$ (BFO) exhibits the steepest switching behavior, making it particularly promising candidate for extremely low power applications.

**Supporting Information 9. Output characteristics of NC-TIFETs**

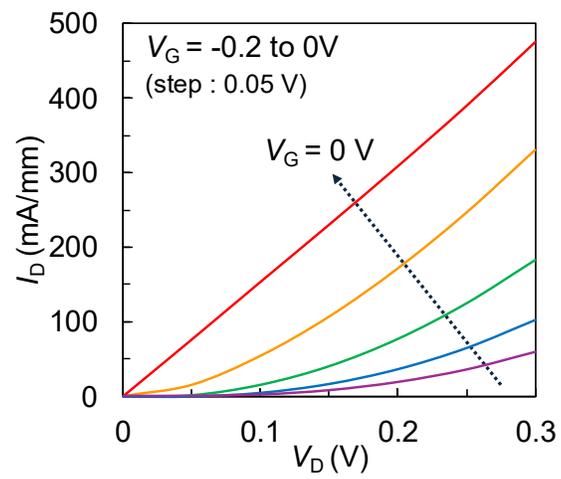

**Figure S7.** Output characteristics of NC-TIFETs with different $V_G$. Ferroelectric material is HZO and channel material is 1T′-MoS$_2$.

**Supporting Information 10. Formulas for TIFET drain current equation combined with contact resistance.**

In this study, the TIFET was modeled as a two-terminal device, where the drain current ($I_D$) is determined by the series combination of two primary resistance components: the contact resistance ($R_C$) at the source and drain, and the intrinsic channel resistance ($R_{CH}$). The total resistance of the device ($R_{tot}$) is the sum of these components (i.e., $R_{tot} = 2R_C + R_{CH}$)[26]. Therefore, the total resistance of the TIFET was modeled as $R_{tot} = 2R_C + R_{CH} = 2R_C + 1/G_0$, where $G_0 = 2e^2/h$ is the quantum conductance. Furthermore, to incorporate the voltage drop induced by the $R_C$, intrinsic supply voltage ($V_D'$) is calculated as $V_D' = V_D - 2R_C I_D$[27], where $V_D$ is the supply voltage.

According to Ohm's law, the $I_D$ is the $V_D'$ divided by the $R_{tot}$:

$$I_D' = \frac{V_D'}{R_{tot}} = \frac{V_D - 2R_C I_D}{2R_C + 1/G_0} \tag{9}$$

To solve for $I_D$, equation (9) is expressed as equation (10):

$$I_D' = \frac{V_D}{4R_C + 1/G_0} \tag{10}$$

In the case of gate-dependent transport in TIFETs, where the channel conductance $G(V_G) = I_D(V_G)/V_D$ is a function of gate bias, the expression for the drain current with contact resistance is given by equation (11):

$$I_D'' = \frac{V_D}{4R_C + \dfrac{1}{G(V_G)}} \tag{11}$$

The transfer characteristics calculated using equation (11) are illustrated in Figure S8.

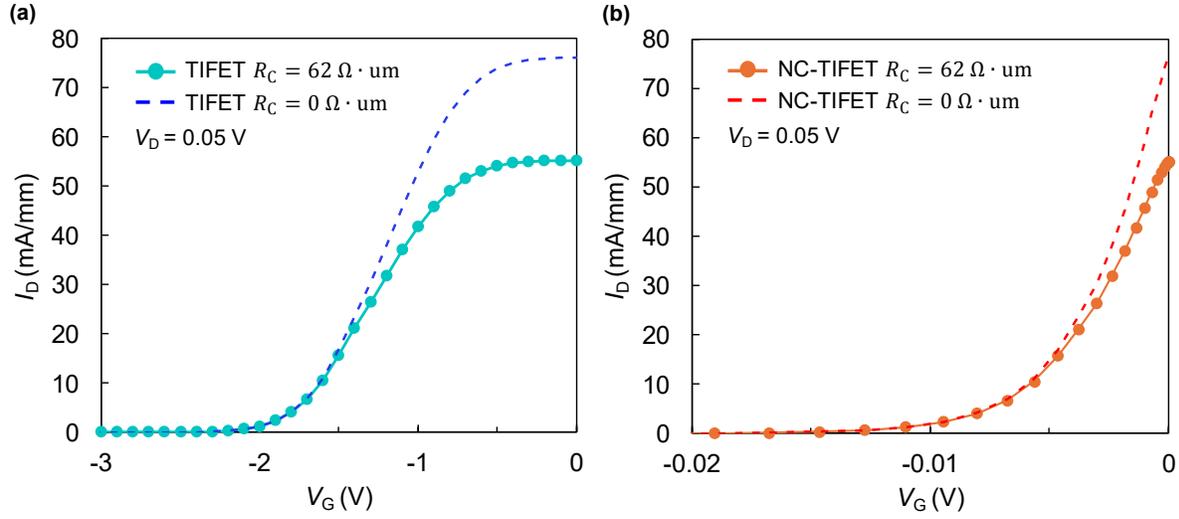

**Figure S8.** Impact of contact resistance ($R_C$) on the transfer characteristics of (a) TIFET and (b) NC-TIFET at $V_D = 0.05$ V. The dashed lines represent the ideal case without contact resistance ($R_C = 0$ Ω·μm), while the solid lines with symbols represent the realistic case with an optimized contact resistance of $R_C = 62$ Ω·μm, as reported $R_C$ for InSe contacts[27]. Note that the TIFET employs a $SiO_2$ dielectric with a 1T′-$MoS_2$ channel, whereas the NC-TIFET utilizes an HZO ferroelectric with a 1T′-$MoS_2$ channel.

Note that contact resistance ($R_c = 62$ Ω·μm) is adopted from experimental demonstrations of ohmic contacts[25]. This assumption of ohmic contact is critical for minimizing contact resistance at cryogenic temperatures where thermionic emission is suppressed. Quantitatively, for the simulated device width of $W = 50$nm, the total series contact resistance is estimated to be $\frac{2R_C}{W} \approx 2.48$ kΩ. This value is notably smaller than the quantized resistance of the 1D helical edge states ($R_{edge} = \frac{h}{2e^2} \sim 12.9$ kΩ). This comparison confirms that the total device resistance is dominated by the intrinsic topological edge channels rather than the contacts, preserving the ballistic transport characteristics essential for the proposed NC-TIFET operation.

As shown in Figure S8(a), the inclusion of $R_C$ leads to a noticeable degradation in the On-current for standard TIFETs. However, for NC-TIFETs, as shown in Figure S8(b), even when accounting for a realistic $R_C$ = 62 Ω·μm, the NC effect effectively preserves the steep switching slope. Although the On-current is slightly reduced compared to the ideal case, the voltage gain and switching capability remain robust, confirming the device's potential for high-performance applications.